\documentclass[9pt,letterpaper]{maxmlr}

\usepackage[
    backend=biber,
    style=authoryear,
    sorting=nty,
    maxbibnames=99,
    natbib=true,
    giveninits=true,
    uniquename=init,
    doi=true,
    url=true,
    isbn=true
]{biblatex}
\definecolor{mydarkblue}{rgb}{0.102,0.451,0.910}
\usepackage[
    pdfauthor={Antonio Max},
    pdftitle={High vs. Low AGI: Ontology and Conceptual Taxonomy for Geopolitical Coherence},
    pdfsubject={Geopolitical Coherence},
    pdfkeywords={AGI Taxonomy, Dual--Use, Securitization Theory, High/Low Politics, Construal-level Theory},
    colorlinks=true,
    linkcolor=mydarkblue,
    citecolor=mydarkblue,
    filecolor=mydarkblue,
    urlcolor=mydarkblue,
]{hyperref}
\usepackage[noabbrev,nameinlink,capitalise]{cleveref}

\DeclareFieldFormat{citehyperref}{%
  \DeclareFieldAlias{bibhyperref}{noformat}% Avoid nested links
  \bibhyperref{#1}%
}
\savebibmacro{cite}
\renewbibmacro*{cite}{%
  \printtext[citehyperref]{%
    \restorebibmacro{cite}%
    \usebibmacro{cite}%
  }%
}

\title{High vs. Low AGI: Ontology and Conceptual\\ Taxonomy for Geopolitical Coherence}
\author{Antonio Max \thanks{Principal Researcher; antoniomaxai@proton.me}}

\begin{document}

\maketitle

\begin{abstract}
    \noindent
    The rapid progression of Artificial General Intelligence (AGI) research demands conceptual tools capable of distinguishing between systems developed for open, commercial integration and those destined for sovereign, securitized deployments. Without such distinctions, risk assessments and regulatory debates collapse AGI into legacy dual-use frameworks that are ill-suited for these resources, capturing the possibility of civilian and military application but overlooking the distinct societal lineages yielded by corporate and state-grade architectures. This paper proposes a taxonomy distinguishing low-AGI and high-AGI, clarifying how commercial-economic and security-sovereign architectures can be distinguished not only by function, but by the social and political ecosystems that produce them. The taxonomy builds on international relations concepts of \enquote{high/low politics,} viewed through the lens of construal-level theory, which allows it to even capture how cooperation and conflict may coexist in the context of AGI's emerging geopolitical stakes. By embedding AGI within power structures and securitization theory, this contribution extends dual-use discourse through an ontological taxonomy that enables more granular risk assessment and governance design—equipping policymakers and researchers to anticipate security dilemmas, institutional demands, and technical-political spillovers in the international system.
\end{abstract}


\section{Introduction}
Artificial General Intelligence (AGI) represents a pivotal advancement in computational capability, enabling systems to perform across heterogeneous domains at or beyond human proficiency. As the technological readiness of AGI accelerates \parencite{feng2024faragillmsneed}, its geopolitical implications demand analytical tools capable of distinguishing civilian-commercial from  sovereign-securitized trajectories of development and deployment.

This paper advances a taxonomy of low-AGI and high-AGI—first outlined in \parencite{Max2025}—to provide a more granular framework for understanding how AGI architectures diverge not only by application but by institutional origin, governance logic, and strategic purpose. Low-AGI captures commercial, public-facing systems optimized for broad economic integration, while high-AGI denotes securitized, state-grade architectures tailored for defense, intelligence, and state survival imperatives. Existing discourse often invokes the phatic notion of \enquote{dual-use} to capture the civilian/military overlap; yet this framing implies a shared technological substrate with different applications, obscuring how strategic actors actively contest, securitize, and reclassify AI capabilities across domains. To overcome this conceptual limitation, the present framework draws from the high/low politics tradition in international relations (IR), which more effectively captures asymmetries and mutual dependencies between economic and security spheres.

Building on \parencite{casier2020not} read of high/low politics, the taxonomy situates AGI development within its broader institutional and philosophical contexts while addressing the interplay of cooperation and competition between low- and high-AGI. It examines the unique challenges of each class and anticipates the nature of spillovers between them, thereby providing a more coherent account of AGI's role in shaping power distribution. Deeper inquiries into how this technological resource translates into political capital—and how its weaponized interdependence unsettles conventional mercantilist logics—should benefit from the ontology developed here. By extending technical classifications associated with AGI competence (e.g., \parencite{morris2023levels}), this work embeds AGI within power structures, ethical considerations, and geopolitical dynamics, empowering policymakers and researchers with a conceptual scaffold for interpreting AI rivalries, refining strategies for digital sovereignty, and developing a governance language more adequate to the demands of high-stakes technological competition \parencite{uscc2024recommendations}.

\section{Theoretical Foundations: High/Low Politics in IR Applied to AGI}

The distinction between high and low politics constitutes one of the most enduring analytical frameworks in International Relations scholarship. Classical realist formulations positioned high politics as matters of national security, survival, and competition—the domain of states pursuing power in an anarchic system—while low politics encompassed economic cooperation, technical coordination, and welfare issues deemed secondary to existential concerns \parencite{morgenthau1948politics,waltz1979theory}\footnotemark, a hierarchical ordering that reflected realism's state-centric ontology and its prioritization of military-strategic considerations. Liberal institutionalists subsequently challenged both the strict hierarchy and the autonomy assumption, arguing that complex interdependence creates linkages across issue areas, with cooperation in low politics domains potentially spilling over to transform high politics dynamics \parencite{keohane2012power}. The high/low distinction thus became contested terrain, with scholars debating not only the boundary classification but whether the distinction itself retained analytical utility in an era of globalization and transnational networks \parencite{ripsman2005false}.
\footnotetext{Waltz (1979, p. 94) cautions that \enquote{the distinction frequently drawn between matters of high and low politics is misplaced,} noting that states often leverage economic instruments for political and military ends and vice versa. This underscores that while realist theory privileges security concerns, economics operates as an instrumental component of power politics rather than a subordinate sphere, as maintained by institutional liberals.}

\cite{casier2020not} intervention reconceptualizes this framework by shifting from ontological claims about issue hierarchies to a constructivist understanding of how political actors themselves categorize classes as vital or routine. His analysis of EU-Russia relations demonstrates that the high/low distinction operates not as an objective property of issue class but as a \enquote{subjective categorization in political discourse itself}\citep[4]{casier2020not}. What matters analytically is how policymakers frame certain areas as touching core national interests requiring paradigmatic positioning (\textit{high politics}) versus domains characterized by multi-actor structures, diffuse interests, and pragmatic interaction (\textit{low politics}). Critically, Casier emphasizes the relative autonomy of different domains: trade, energy, and educational exchanges can maintain cooperative patterns even as security relations deteriorate into confrontation. This autonomy challenges liberal expectations of functional spillover while avoiding realist assumptions of strict hierarchy: individual \enquote{geopolitical fields} within low politics possess their own logics and can display varying patterns of cooperation and conflict independent of broader strategic tensions.

Casier grounds this reconceptualization in Construal-Level Theory (CLT), a social psychological framework that explains how mental representations vary with psychological distance \parencite{trope2010construal}. CLT posits that objects, events, or actors psychologically distant from direct experience—whether temporally, spatially, socially, or hypothetically—are construed abstractly, reduced to essentialized characteristics and governed by ideological schemas. Proximate phenomena, conversely, are construed concretely, with attention to contextual specifics and immediate particularities. Applied to international relations, low politics involves small psychological distances: actors interact directly, frequently, and in specific contexts (negotiating contracts, solving technical problems, coordinating logistics), their mental representations of counterparts remain concrete, grounded in immediate experience rather than abstract attributions. High politics, by contrast, operates at high psychological distance: political leaders interpret adversary behavior through ideologized frameworks, attribute intentions based on essentialized images rather than direct interaction, and engage in paradigmatic positioning divorced from the daily reality of interaction \citep[6]{casier2020not}. Importantly, CLT predicts asymmetric contagion: concrete construals do not easily penetrate abstract ones, meaning cooperative practices in low politics domains rarely transform polarized discourses in high politics, while top-down imposition of conflict (through sanctions, for instance) requires deliberate political decisions rather than spontaneous spillover.

This theoretical architecture translates productively to AGI governance. The high/low-AGI ontology proposed here adapts Casier's framework across four dimensions: (1) actor structure—whether governance occurs through dispersed technical communities and commercial entities (low-AGI) or concentrated among state security establishments and elite decision-makers (high-AGI); (2) psychological distance and abstraction—whether AGI developments are construed concretely as specific technical challenges with immediate, localized implications, or abstractly as existential risks requiring ideological positioning around civilizational values; (3) discursive framing—whether AGI issues are debated in pragmatic, domain-specific terms or elevated to grand strategic narratives about technological supremacy and regime survival; and (4) spillover dynamics—whether progress in technical standardization, safety research, or commercial cooperation can transform geopolitical competition, or whether strategic distrust forecloses functional cooperation. Casier's demonstration that EU-Russia trade and energy cooperation persisted despite profound strategic crisis suggests that AGI governance may similarly fragment: technical cooperation on specific assets or transparency measures may coexist indefinitely with great power competition over AGI development timelines and deployment doctrines. The relative autonomy of governance domains, rather than integrated cooperation or comprehensive conflict, may characterize the emerging international AGI order.

\section{The High/Low AGI Dichotomy: Dissecting the Archetypes}

Although existing research in AI geopolitics frequently frames artificial intelligence through conventional dual-use lenses—civilian versus military applications, or commercial versus sovereign deployments—such framings remain conceptually inadequate for analyzing the strategic differentiation emerging within advanced AI development. The dual-use concept, borrowed from export control regimes and nonproliferation frameworks, treats technology as having a primary legitimate purpose that is good or at least non-harmful, with harmful consequences arising from unintended secondary applications \parencite{forge2010note}. This occludes how AI systems engineered for commercial deployment often embed architectural logics optimized for scalability, user accountability, and rapid iteration, in stark contrast to those designed for strategic state purposes, which inherently prioritize operational resilience, opacity, and sovereign control as \textit{intended} primary functions—even when both draw on similar foundational techniques that lack primary or secondary harmful intentions, until securitized by actors as existential threats that collapse the taxonomy altogether. 

This work advances beyond binary dual-use categorization by developing a granular ontology of AGI grounded in \parencite{casier2020not} reconceptualization of high and low politics. As established in Section 2, Casier demonstrates that the high/low distinction operates not as objective hierarchy but as \textit{constructed categorization} reflecting how actors frame domains as touching vital interests versus routine interaction. Crucially, however, these framings do not float free from material conditions: certain structural features make domains more amenable to treatment as low politics (multi-actor complexity, diffuse interests, frequent direct interaction) while others invite high politics framing (concentrated decision-making, paradigmatic positioning, abstract threat construals). \textbf{Table} \ref{tab:agi_ontologies} operationalizes this insight by identifying the technical, institutional, and normative characteristics that emerge when AGI development is framed and organized as low-politics versus high-politics domains, while also examining which structural features make such framings more or less sustainable over time.

\begin{table}[!ht]
    \vspace{0.5em}
    \small
    \caption[\hspace{0pt}]{\raggedright Ontological Archetypes of Sovereign AGI R\&I and Dual-use Powers\par}
    \vspace{-0.7em}
    \centering
    \label{tab:agi_ontologies}
    \begin{tabularx}{\textwidth}{p{0.13\textwidth} XX}
    \toprule
    \textbf{} & \textbf{Public-facing (Low) AGI} & \textbf{State-grade (High) AGI} \\
    \midrule
    \textbf{Character} & Commercial—AGI as a product or service. & Sovereign—AGI as a lever of power. \\
    \midrule
    \textbf{Nature} & A tool for augmentation and production. Its core purpose is integration into economic structures to increase efficiency, provide convenience, and generate profit. It is a utility, a platform, a consumer product. & A strategic asset. Its core purpose is to project power, achieve military and intelligence objectives, and ensure national security. It is fundamentally a tool for survival and dominance. \\
    \midrule
    \textbf{Architecture} & Designed for scale and broad applicability. It is probabilistic and open to a wide range of inputs. Its mindset is one of generalized problem-solving within civilian, regulated, and legally circumscribed frameworks. Data is readily available via commercial means. & Designed for robustness, security, and specific effectiveness. Leans toward deterministic, composable closed-loop systems, a force multiplier under national security mandates, with a goal to deny and control, not only augment. Secure data is scarce. \\
    \midrule
    \textbf{Ethos} & Guided by market demand, shareholder value, and a public-facing ethical framework. Its being is defined by its ability to serve users and be a productive member of a capitalist economy. It must be seen as benevolent. & Guided by a strategic calculus of deterrence and preemption. Its ethical framework is internal and tied to military necessity and national interest in zero-sum competition. Its being is defined by its adaptive ability to outperform an adversary's system. \\
    \midrule
    \textbf{Governance} & Shaped by external oversight (regulators, public opinion, social status). Accountability flows outward. Conservative on reputational and liability risks. & Shaped by administrative secrecy and national command structures. Accountability flows upward to the state. Higher risk tolerance for critical applications. \\
    \midrule
    \textbf{Tempo} & Operates on quarterly earning cycles and product release schedules, optimizes for near-term market capture and user retention. & Continuous development, intermittent deployment, mission-oriented life-cycles. Responds to geopolitical time rather than market patterns. \\
    \bottomrule
    \end{tabularx}
    \par\vspace{0.5em}
    \footnotesize{Source: Adapted from \parencite{Max2025}.}
\end{table}

Low-AGI encompasses public-facing systems characterized by commercial imperatives, broad applicability, and external accountability. These systems exhibit the structural features Casier associates with low politics: development involves diffuse actor networks (researchers, engineers, product managers, users), interactions occur at low psychological distance (solving specific technical problems, optimizing particular applications), and governance operates through concrete regulatory frameworks rather than abstract strategic doctrines typical of state-grade deployments. Foundation models trained on public datasets and deployed as consumer services exemplify this category—their development timeline responds to market signals and quarterly earnings cycles rather than geopolitical events, and their architectures prioritize generalizability over mission-specific robustness \parencite{DBLP:journals/corr/abs-2108-07258}.

High-AGI, conversely, comprises state-grade systems positioned as sovereign capabilities for intelligence, military, and strategic applications. These systems mirror Casier's high politics characteristics: development concentrates among security establishments and elite decision-makers, psychological distance from direct experience increases (designers reason about abstract adversary intentions rather than concrete user needs), and governance operates through classified command structures emphasizing operational security over public accountability. Architectures prioritize deterministic reliability and adversarial robustness over generalized capability, reflecting what CLT predicts for high psychological distance: construals become abstract, essentialized around existential concerns (deterrence, preemption, strategic advantage) rather than contextual specifics \parencite{trope2010construal}. Development tempo aligns with mission cycles and threat assessments rather than product roadmaps, and ethical frameworks internalize within national security logic rather than external stakeholder consultation \parencite{unidir2025ai,allen2017artificial}.

The architectural dimensions warrant emphasis. Low-AGI systems adopt probabilistic, open-ended designs optimized for scale and adaptability across diverse civilian contexts—what Casier would characterize as concrete construals responsive to immediate, varied use cases. High-AGI systems increasingly favor compositional, closed-loop architectures: modular components (specialized sensors, reasoning engines, action modules) integrated under deterministic control protocols \parencite{baeza2025aidriventacticalcommunicationsnetworking,DoD-JWSWG-2024,nato2024ai,eda2025trustworthiness}. This mirrors biosafety level gradations where increasing hazard drives architectural closure and procedural formalization \parencite{pannu2025dual}. The ethos contrast reflects CLT's prediction: low psychological distance (commercial AGI developers interacting with users, regulators, media) produces concrete ethical reasoning around specific harms and benefits; high psychological distance (strategic AGI planners reasoning about abstract adversary capabilities) produces ideologized frameworks centered on national survival and zero-sum competition \parencite{mearsheimer2014tragedy}.

Critically, this ontology unpacks AGI's profound hybridity and the contagion dynamics Casier identifies as central to understanding cooperation-conflict coexistence. Low-AGI innovations— breakthrough architectures, training techniques, scaling laws—do not remain confined to commercial domains: talent migration, reverse-engineering, and espionage enable high-AGI programs to adapt civilian advances. This technical diffusion, however, should not be confused with the political contagion Casier examines. The flow of knowledge artifacts (algorithms, libraries, repositories) across the low/high boundary does not carry with it the cooperative practices, trust-building interactions, or concrete construals that characterize low-AGI development. High-AGI programs extract innovations from open and commercial ecosystems while maintaining the abstract threat construals, zero-sum logics, and securitized governance structures characteristic of high politics. The EU's continued reliance on Russian natural gas did not transform Moscow's strategic calculations about NATO expansion; similarly, China's absorption of open-source AI research does not mitigate Washington's threat perceptions about AGI-enabled surveillance or autonomous weapons. Technical interdependence absent institutional cooperation may even exacerbate security dilemmas, as \parencite{krickovic2015interdependence} demonstrated for energy relations.

The inverse contagion—from high to low politics—also warrants attention. Casier demonstrates that conflictual dynamics trickle down primarily through deliberate political decisions (e.g., sanctions) rather than spontaneous spillover, because abstract construals do not automatically contaminate concrete ones. Applied to AGI: export controls, investment restrictions, and talent visa policies represent top-down impositions of strategic logic onto commercial domains \parencite{uscc2024hearing}. The critical question is whether such interventions successfully convert low-AGI fields into high-AGI domains, or whether the multi-actor complexity and diffuse interests of commercial AI maintain relative autonomy despite political pressure—as seen in the U.S. framing of European reliance on Chinese AI hardware as a security threat, an attempt to elevate commercial AI into high politics whose prospects remain uncertain given entrenched economic interdependencies \parencite{arcesati2023aientanglements}. This represents what securitization theory terms a \enquote{speech act}—political elites performatively constructing commercial AI relationships as existential threats requiring emergency measures beyond normal regulatory channels \parencite{langenohl2019conceptualizing}. Whether such securitization succeeds depends on audience acceptance: do European firms, publics, and policymakers accept the threat framing, or do entrenched commercial interests and concrete positive experiences with Chinese technology resist the abstract security narrative? The outcome remains contested precisely because high psychological distance construals compete with low psychological distance experiences.

This dissection establishes that AGI governance cannot be understood through unified frameworks assuming either comprehensive cooperation or totalizing competition. Instead, the high/low ontology predicts fragmented governance: technical standards and safety protocols may advance through low-AGI channels characterized by pragmatic multi-stakeholder collaboration, while strategic AGI development proceeds through high-AGI channels marked by secrecy, distrust, and arms race dynamics. The relative autonomy of these domains, their asymmetric contagion patterns, and the ongoing contestation over which AGI developments belong in which category constitutes the core \textit{problematique} for international AGI governance in an era of great power competition.

\section{Ontological Implications for the Geopolitics of AGI}

The high/low-AGI ontology developed above generates three critical implications for understanding the emerging geopolitical order around artificial general intelligence. First, it predicts governance fragmentation rather than regime convergence or comprehensive breakdown. Second, it unpacks how securitization dynamics operate differently across AGI domains, with varying prospects for success. Third, it reveals the strategic significance of boundary contestation—struggles over which AGI developments belong in which ontological category become themselves sites of geopolitical competition.

\subsection{Fragmented Governance and the Limits of Regime Complexity}

Conventional analyses of global technology governance often invoke regime complex theory to explain institutional proliferation and regulatory overlap \parencite{raustiala2004regime,orsini2013regime}. Emerging from efforts to make sense of fragmented global governance, regime complexes—loosely coupled sets of institutions addressing a single issue domain through partially overlapping rules and norms—have been used to describe areas from climate change to global health and intellectual property. Applied to AI governance, this framework anticipates a patchwork of multilateral forums (UN, OECD), plurilateral initiatives (Partnership on AI, Global Partnership on AI), bilateral agreements, and multistakeholder arrangements spanning the public-private divide \parencite{cihon2020fragmentation}. Yet the high/low-AGI ontology suggests that this framework may misdiagnose the fundamental structure of AGI governance: regime complexes presume that diverse institutional arrangements govern a \textit{single issue domain} from different angles—an architecture prone to forum shopping, inconsistent implementation, and strategic ambiguity \parencite{keohane2011regime}—whereas the high/low distinction shows AGI governance does not constitute a single domain but rather \textit{two} partially decoupled spheres operating under fundamentally different logics.

Low-AGI governance resembles genuine regime complexity: overlapping institutions (IEEE standards bodies, EU AI Act, corporate AI ethics boards, NIST AI Risk Management Framework) address commercial AI deployment through multi-stakeholder deliberation, technical standardization, and iterative regulatory refinement. Actors navigate institutional multiplicity strategically, but share basic commitments to transparency, accountability frameworks, and balancing innovation with risk mitigation. Cooperation occurs through low psychological distance interaction—regulators meet with engineers, civil society organizations audit deployed systems, researchers publish benchmark evaluations. The regime complex framework captures this domain reasonably well.

High-AGI governance, conversely, escapes regime complexity altogether. Strategic AI development occurs through \textit{dualistic structures} rather than overlapping regimes: NATO versus Chinese military AI programs, Five Eyes intelligence sharing versus Russian cyber capabilities, U.S. semiconductor export controls versus Chinese self-sufficiency campaigns. These are not multiple institutions governing a shared problem but opposed security architectures pursuing mutually exclusive objectives. No meaningful regime exists because the preconditions for institutionalized cooperation—shared normative frameworks, transparency about capabilities and intentions, verification mechanisms—are absent \parencite{horowitz2021speed}. Casier's EU-Russia analysis demonstrates that such dualistic structures \textit{entrench} polarization rather than managing it: \enquote{This type of dualistic structure tends to stimulate highly abstract images of the adversary and his intentions rather than mitigating them} \citep[7]{casier2020not}.

The implications are stark. Efforts to build comprehensive AGI governance regimes spanning commercial and strategic development—proposals for global AGI safety summits, international compute governance, automatic monitoring, or universal AI development moratoria—confront structural obstacles beyond diplomatic will or institutional design \parencite{shavit2023practices,trager2023international}. Low-AGI domains may sustain fragmented but functional cooperation even amid broader geopolitical competition, as exemplified by how semiconductors, cloud services or data flows remain largely commercial but are increasingly securitized via imperatives of autonomy and sovereignty \parencite{draghi2024future}. High-AGI domains, however, resist overt institutionalization because the psychological distance separating security establishments, the abstract threat construals governing strategic planning, and the absence of direct interaction between adversary programs foreclose the trust-building necessary for regime formation. The governance fragmentation this produces is not a coordination failure amenable to better institutional architecture but a structural feature of ontologically distinct domains operating under incommensurable logics.

\subsection{Differential Securitization: When and Why Commercial AI Becomes Strategic}

The relative autonomy of low and high-AGI domains does not imply impermeable boundaries. As Section 3 established, deliberate political decisions can impose high politics logic onto low politics fields through what securitization theory conceptualizes as performative speech acts. Understanding when and why such securitization succeeds or fails reveals a key geopolitical dynamic: the political struggle to redefine specific AGI domains as existential threats requiring extraordinary measures.

Securitization theory, developed by the Copenhagen School, argues security is not an objective condition, but a social construction achieved through discourse \parencite{langenohl2019conceptualizing}. A \enquote{securitizing move} occurs when political actors (securitizing agents) designate something as an existential threat to a referent object (the nation, regime survival, civilizational values), claiming the threat's urgency justifies emergency measures transcending normal political procedures. Successful securitization requires not only speech acts by credible elites but audience acceptance—publics, legislatures, or relevant stakeholders must accept the threat framing and tolerate extraordinary responses \parencite{stritzel2007towards}.

Applied to AGI, the high/low ontology predicts differential securitization success across domains. Low-AGI fields resist securitization because actors' concrete construals, derived from direct experience and low psychological distance interaction, compete with abstract threat narratives. Consider foundation model development: researchers interact daily with these systems, understanding their capabilities, limitations, and failure modes through immediate experimentation. When security elites frame foundation models as civilizational threats requiring development prohibitions, this abstract construal confronts the concrete knowledge of practitioners who see incremental capability gains, manageable risks through technical interventions, and substantial societal benefits. The gap between abstract securitizing discourse and concrete professional experience creates friction limiting securitization success \parencite{hindriks2023risks}.

High-AGI domains, conversely, exhibit high psychological distance: few actors possess direct experience with adversary strategic AI systems, capabilities remain classified, and intentions must be inferred from abstract geopolitical positioning rather than concrete interaction. This makes securitization easier—absent contradictory direct experience, abstract threat construals face minimal resistance. When U.S. officials frame Chinese military AI as enabling \enquote{AI-powered authoritarianism} threatening liberal democracy, or Chinese scholars describe American AGI programs as instruments of \enquote{technological hegemony} \parencite{ding2018deciphering,roberts2021chinese}, these construals operate in low-information environments where psychological distance prevents concrete alternative framings from emerging. Securitization succeeds because it constructs the terrain on which high-AGI competition occurs rather than competing with existing non-securitized understandings.

The critical geopolitical question becomes: can securitizing actors successfully convert low-AGI domains into high-AGI domains by elevating commercial AI development into existential threat narratives? Three factors condition success:

\textbf{First, technological trajectory matters.} If AGI capabilities advance toward systems plausibly enabling decisive military advantages—autonomous strategic planning, unhackable cyber operations, actionable intelligence analysis—the technical reality increasingly validates security framings regardless of actors' concrete experiences. The gap between current commercial AI (useful but limited tools) and hypothetical transformative AGI (civilizational-scale power) creates space for securitization as capabilities approach the latter threshold \parencite{bostrom2014superintelligence,ord2020precipice}.

\textbf{Second, institutional access determines whose construals dominate.} Security establishments possess unique capacities to classify information, restrict participation in strategic AI programs, and control narrative framing through classification regimes. By moving AGI development behind security perimeters, states \textit{create} the high psychological distance that enables abstract threat construals to dominate \parencite{bullock2025agigovernmentsfreesocieties}. If commercial AI development remains predominantly in low-AGI registers—published research, open-source implementations, multi-stakeholder governance—concrete construals resist securitization. If states successfully channel advanced development into classified programs, they engineer the conditions for securitization's success.

\textbf{Third, geopolitical context modulates receptivity.} Broader strategic competition primes audiences to accept securitization across domains. The U.S.-China \enquote{strategic competition} framing creates receptive conditions for securitizing commercial AI relationships: export controls on semiconductors, restrictions on Chinese AI investment in U.S. firms, and visa limitations on Chinese AI researchers all reflect successful securitization enabled by ambient threat perceptions \parencite{allen2022choking}. Conversely, the EU's relative geopolitical autonomy and economic interdependence with China creates domestic audiences more resistant to securitizing commercial AI—European firms' concrete positive experiences with Chinese markets compete with transatlantic security narratives \parencite{draghi2024b}.

The implication: AGI geopolitics increasingly centers on \textit{boundary struggles}—efforts to move particular AGI developments from low-AGI registers (commercial, collaborative, transparently governed) into high-AGI registers (securitized, classified, zero-sum). These are not technical classifications but political projects with profound consequences for governance possibilities.

\subsection{Boundary Contestation as Geopolitical Strategy}

If the high/low-AGI boundary is constructed rather than inherent, its location becomes itself a site of geopolitical competition. Actors pursue boundary strategies—deliberately attempting to move AGI domains across the high/low divide—to advance strategic objectives. Three such strategies warrant attention.

\textbf{Securitization offensives} involve state actors attempting to reclassify commercial AI domains as national security threats requiring state control. The United States' evolving approach to semiconductor technology illustrates this dynamic: formerly governed through commercial export licensing (low politics), cutting-edge chip production now falls under national security restrictions justified by military-AI implications (high politics). The October 2022 export controls targeting Chinese access to advanced semiconductors and chip-making equipment represent successful securitization—converting commercial supply chains into strategic choke points through threat construction \parencite{khan2021semiconductor,miller2022chip}. This strategy serves dual functions: directly constraining adversary capabilities while establishing precedents for treating AI-enabling technologies as inherently strategic rather than commercial.

\textbf{De-securitization campaigns} pursue the opposite: resisting or reversing securitization to maintain commercial AI development in low-AGI registers and European responses to U.S. pressure for decoupling from Chinese AI supply chains exemplify this strategy. By emphasizing concrete economic costs, technical interdependencies, and the absence of direct security threats from commercial AI collaboration, European policymakers resist American securitizing framings \parencite{arcesati2023aientanglements}. Academic and industry communities similarly pursue de-securitization through open publication norms, international research collaborations, and emphasis on globally shared AI risks (e.g., climate modeling, pandemic preparedness) requiring cooperative rather than competitive approaches \parencite{AIOpenness2025}. These efforts combat the psychological distance escalation that enables abstract threat construals by maintaining concrete collaborative practices.

\textbf{Strategic ambiguity exploitation} involves maintaining AGI programs that straddle the high/low boundary, enabling actors to access low-AGI resources (talent, data, compute, algorithms) while developing high-AGI capabilities. China's \enquote{military-civil fusion} strategy exemplifies this approach: nominally commercial AI firms with deep ties to military and intelligence establishments develop dual-use capabilities, accessing global supply chains and research networks (low-AGI register) while channeling advances to strategic applications (high-AGI register) \parencite{kania2021seizing}. This strategic ambiguity frustrates governance efforts—should commercial partnerships with Alibaba or Tencent be treated as low-AGI cooperation or high-AGI security threats? The ambiguity itself becomes strategically valuable, enabling resource access that would be foreclosed by clear high-AGI categorization.

These boundary strategies generate recursive dynamics. Securitization offensives by one actor prompt de-securitization resistance by others, while strategic ambiguity exploitation triggers counter-securitization efforts to \enquote{unmask} dual-use programs. The U.S.-China technology competition increasingly centers on these boundary struggles: American export controls attempt to securitize semiconductors and AI training infrastructure; Chinese industrial policy attempts to de-securitize access to global technology supply chains; and both powers accuse the other of strategic ambiguity regarding military AI development \parencite{fedasiuk2021harnessed}.

The geopolitical implication extends beyond bilateral competition, as boundary location determines governance possibilities: domains remaining in low-AGI registers can sustain multilateral cooperation, technical standardization, and verification mechanisms; domains successfully securitized into high-AGI registers foreclose such cooperation. The progressive securitization of AGI-relevant domains therefore constitutes an endogenous threat construction: framing AGI as existential strategic competition creates the conditions (secrecy, distrust, arms race dynamics) that validate the framing while foreclosing cooperative alternatives \parencite{barnhart2022emerging,dafoe2024ai,chessen_martell_2025}. Yet, as Casier's framework predicts, complete securitization across all AGI domains remains unlikely—the multi-actor complexity, concrete interactions, and diffuse interests characterizing low-AGI fields provide structural resistance to comprehensive securitization, even as high-AGI domains spiral into unmanaged competition.

\subsection{Synthesis: Toward Fragmented Coexistence}

The ontological analysis yields a counterintuitive geopolitical prediction: the emerging AGI order will likely resemble neither the cooperative governance regimes envisioned by multilateralist optimists nor the unrestrained strategic competition feared by security pessimists. Instead, a \textit{fragmented coexistence} regime becomes probable—indefinite persistence of cooperation in some AGI domains alongside unmanaged competition in others, with limited contagion between spheres.

This fragmentation reflects the structural features the high/low-AGI ontology exposes: psychological distance differentials across domains, asymmetric contagion dynamics, and boundary contestation. Low-AGI governance may achieve modest successes—international standards for AI transparency, coordinated approaches to algorithmic bias, even limited verification mechanisms for commercial systems—not because strategic competition abates but because these domains maintain relative autonomy from high politics. High-AGI competition may simultaneously intensify—arms racing in military AI, espionage targeting strategic programs, zero-sum positioning over AGI timelines—without destroying low-AGI cooperation, because abstract threat construals at high psychological distance do not automatically contaminate concrete collaborative practices at low psychological distance.

The EU-Russia analogy again proves instructive: six years into profound strategic crisis, energy cooperation persists, trade has recovered, and academic exchanges continue at high levels, yet the strategic partnership remains suspended and sanctions renewed regularly. This is not transitional instability awaiting resolution toward either renewed partnership or complete breakdown but rather a stable equilibrium reflecting domain autonomy and asymmetric contagion \parencite{casier2020not}. AGI governance may similarly stabilize into awkward coexistence—cooperative progress on AI safety standards coexisting with secretive AGI weapons development, open-source research communities persisting amid export-controlled compute infrastructure, multilateral AI ethics declarations coinciding with bilateral AGI supremacy competition.

Such fragmentation frustrates both policy communities. Those prioritizing existential risk mitigation confront the reality that technical safety research in low-AGI registers may advance substantially without constraining strategic risk-taking in high-AGI domains. Those prioritizing geopolitical advantage confront the reality that securitization efforts to comprehensively control AGI development face structural limits from the multi-actor complexity and international interdependencies defining low-AGI fields. The geopolitical order emerging around AGI may thus be characterized by productive cooperation in areas of low perceived strategic stakes alongside dangerous competition in areas constructed as existential—not because actors choose suboptimal outcomes but because the ontological structure of AGI development sustains both simultaneously.

\section{Conclusion}

This paper extends existing AI governance frameworks beyond conventional dual-use classifications by introducing a structural high/low-AGI framework derived from construal-level theory and the concept of high/low-politics (Casier, 2020). It identifies structural and cognitive variables—such as psychological distance, actor configuration, and governance logic—that differentiate commercial, public-facing AI systems from sovereign, state-grade AGI programs—distinctions obscured by dual-use classifications that presume a primary good or non-harmful purpose, while overlooking how strategic deployments can embed harmful or zero-sum functions from inception—functions that only become formally recognized as existential threats once securitized.

Three core findings emerge. First, AGI governance is likely to stabilize into a multi-domain equilibrium, where cooperative and competitive dynamics coexist across different capability levels and institutional contexts. Pragmatic cooperation in low-AGI domains (technical standards, safety research, commercial deployment frameworks) can persist alongside strategic competition in high-AGI domains (military applications, intelligence systems, deterrence architectures), with asymmetric contagion limiting spillover between spheres. This prediction challenges both multilateralist expectations of functional cooperation transforming strategic relations and realist assumptions that economic interdependence necessarily subordinates to security competition.

Second, the central geopolitical dynamic centers on boundary contestation—struggles over whether specific AGI developments belong in low-AGI or high-AGI registers. State actors pursue securitization offensives to reclassify commercial AI as strategic threats requiring extraordinary controls; commercial and scientific communities resist through de-securitization emphasizing concrete collaborative benefits; and some actors exploit strategic ambiguity to access low-AGI resources while developing high-AGI capabilities. These boundary struggles shape both developmental trajectories and governance possibilities, as domains successfully securitized into high-AGI registers foreclose the multi-stakeholder cooperation sustainable in applied AI and low-AGI fields.

Third, the framework provides operational advantages for granular governance design unavailable through dual-use classifications. By anchoring distinctions in observable structural features—actor configurations, psychological distance in decision-making, accountability flows, operational tempos, and architectural characteristics—the ontology enables systematic tracking of which systems approach strategic thresholds. Unlike dual-use frameworks that conflate diverse AI systems under binary civilian/military categories, the high/low-AGI distinction permits graded responses: harmonized transparency standards for commercial deployment coexisting with verification protocols for strategic systems, or export controls calibrated to architectural characteristics rather than blanket technology categories. Critically, the framework identifies when systems traverse boundaries—when commercial AI development migrates into classified programs, when open research becomes export-controlled, when multi-stakeholder governance yields to security classification—further enabling proactive rather than reactive governance.

For international institutions, this provides a blueprint for developing differentiated regulatory architectures. Rather than pursuing comprehensive AGI governance regimes unlikely to overcome the psychological distance and abstract threat construals characterizing high-AGI competition, multilateral venues can focus efforts where structural conditions support cooperation: technical standards bodies for low-AGI systems, transparency mechanisms for boundary monitoring, and graduated verification protocols sensitive to the legitimate security concerns that make high-AGI programs resistant to intrusive oversight. This approach acknowledges that AGI's capability-continuous development occurs across fundamentally divergent institutional logics—commercial innovation dynamics versus strategic security imperatives—and designs governance accordingly.

However, the ontology's analytical traction depends on continued empirical validation: future research should examine: (1) whether predicted fragmentation patterns emerge as AGI capabilities advance; (2) which securitization attempts succeed in converting low-AGI domains, and why; (3) how boundary location affects international cooperation on specific governance challenges (compute governance, safety standards, verification); and (4) whether construal-level effects on perceived risk and control can be empirically observed in AGI governance decision-making contexts. By grounding the high/low distinction in falsifiable structural claims rather than definitional stipulations, the framework invites empirical scrutiny that can refine or revise its categories.

Ultimately, understanding AGI geopolitics requires moving beyond debates over whether cooperation or competition will dominate toward recognizing that both will coexist in domain-specific patterns. The high/low-AGI ontology provides conceptual architecture for analyzing this coexistence: specifying which structural features sustain cooperation despite strategic competition, identifying how boundary contestation shapes governance possibilities, and enabling policy design calibrated to the divergent logics governing commercial and state-grade systems. As AGI development progresses, the capacity to disambiguate these domains analytically may prove as strategically consequential as the technologies themselves.

\printbibliography
\end{document}